\documentclass[prl,twocolumn,a4paper,superscriptaddress,showpacs,floatfix]{revtex4-1}

\usepackage{dcolumn,amsmath,xspace}
\PassOptionsToPackage{caption=false}{subfig}
\usepackage{subfig}
\usepackage{graphicx}
\usepackage{multirow}

\usepackage[utf8]{inputenc}
\usepackage{lineno}

\usepackage{epstopdf}

\newcommand{\degC}{\ensuremath{~^{\circ}\text{C }}}

\graphicspath{{./img/}}

\begin{document}
\title{Polarization sensitive surface band structure of doped BaTiO$_\mathrm{3}$(001)}

\author{J. E. Rault}
\author{J. Dionot}
\author{C. Mathieu}
\affiliation{CEA, DSM/IRAMIS/SPCSI, F-91191 Gif-sur-Yvette Cedex, France}
\author{V. Feyer}
\affiliation{Peter Gr\"{u}nberg Institute (PGI-6), JARA-FIT, Research Center J\"{u}lich, 52425 J\"{u}lich, Germany}
\affiliation{NanoESCA beamline, Sincrotrone Trieste, Area Science Park, 34149 Basovizza, Trieste, Italy}
\author{C.M. Schneider}
\affiliation{Peter Gr\"{u}nberg Institute (PGI-6), JARA-FIT, Research Center J\"{u}lich, 52425 J\"{u}lich, Germany}
\author{G. Geneste}
\affiliation{CEA, DAM, DIF, F-91297 Arpajon, France}
\author{N. Barrett}
\email[Correspondence should be addressed to ]{nick.barrett@cea.fr}
\affiliation{CEA, DSM/IRAMIS/SPCSI, F-91191 Gif-sur-Yvette Cedex, France}

\begin{abstract}
We present a spatial and wave-vector resolved study of the electronic structure of micron sized ferroelectric domains at the surface of a BaTiO$_\mathrm{3}$(001) single crystal. The n-type doping of the BaTiO$_\mathrm{3}$ is controlled by in-situ vacuum and oxygen annealing, providing experimental evidence of a surface paraelectric-ferroelectric transition below a critical doping level. Real space imaging of photoemission threshold, core level and valence band spectra show contrast due to domain polarization. Reciprocal space imaging of the electronic structure using linearly polarized light provides unambiguous evidence for the presence of both in and out-of plane polarization with two and fourfold symmetry, respectively. The results agree well with first principles calculations.

\end{abstract}
\vspace*{4ex}

\maketitle
\newpage

At a surface or interface of a ferroelectric film the depolarizing field arising from uncompensated surface charges can reduce or even suppress ferroelectricity below a critical thickness.~\cite{Junquera2003} Surface charge can be screened by a variety of mechanisms: intrinsic~\cite{Mi2012, Fong2006}, extrinsic~\cite{Fong2006, Wang2009} or domain ordering.~\cite{Shimada2010, McQuaid2011} Domain ordering and surface atomic structural changes can combine within single domains.~\cite{Meyer2001, Cai2005, Pancotti2013} These charge-driven modifications are likely to strongly influence the surface electronic structure.

Understanding such changes is mandatory for effective integration of ferroelectric materials into oxide-based electronic devices.~\cite{Dawber2005} Any useful device design requires in-depth knowledge of the electronic structure at the interface. For example, interface hybridization between filled d orbitals responsible for magnetization and empty d orbitals in the ferroelectric oxide~\cite{Duan2006} plays a crucial role in heterostructures showing magnetoelectric coupling.~\cite{Ramesh2007} However, studies of the surface properties have focused mainly on polarization, transport and related phenomena.~\cite{Gruverman2002, Kalinin2001}
The anomalous dynamical charge tensors associated with the atomic distortions are particularly high for the oxygen-cation bonding chains.~\cite{Ghosez1998} Measuring the dispersion relations throughout the valence band would therefore be extremely useful.

Angle resolved photoelectron spectroscopy (ARPES) is the ideal technique for measuring the band structure. It is used to study the electronic structure of metallic oxides such as SrRuO$_\mathrm{3}$~\cite{Shai2013} and La$_\mathrm{1-x}$Sr$_\mathrm {x}$MnO$_\mathrm{3}$~\cite{Chikamatsu2007} but rarely applied to insulating perovskite oxides~\cite{King2012}, because of charging during photoemission. Even where a metallic state is expected at the interface between two band insulators such as LaAlO$_\mathrm{3}$ and SrTiO$_\mathrm{3}$ (STO), the substrate is often lightly doped.~\cite{Meevasana2011} Vacuum cleaved stoichiometric STO did allow ARPES measurements but this was due to the appearance of a metallic 2D Fermi surface.~\cite{Santander2011}

A ferroelectric should be an insulator otherwise the long-range Coulomb interactions are screened, suppressing the polarization but there is now evidence that perfect insulators are not necessary to support ferroelectricity. Ferroelectric distortion is predicted and observed in doped BaTiO$_\mathrm{3}$ (BTO) with a critical doping of $1.36-1.9\!\times\!10^{21}$ e/cm$^\mathrm{3}$ for the disappearance of the FE state~\cite{Cochran1960, Hwang2010, Iwazaki2012, Wang2012}. Changes in the unit cell volume induced by oxygen vacancies further reduce the critical doping.~\cite{Iwazaki2012} Thus, within certain limits, a material can have metallic character while stabilizing a ferroelectric state. This opens the way to full characterization of the electronic and chemical structure using photoemission-based experimental techniques.

Micron-scale ferroelectric domain recognition of lightly doped ferroelectric samples can be performed using energy filtered photoelectron emission microscopy (PEEM).~\cite{Barrett2013} PEEM demonstrated the existence of domains in BTO above the Curie temperature due to anionic surface relaxation.~\cite{Hoefer2012}

In this letter we use synchrotron radiation-induced PEEM with linearly polarized light to study the surface band structure of micron sized ferroelectric domains in doped BTO(001). We provide evidence of a surface paraelectric-ferroelectric transition with decreasing metallicity. The microscopic band structure identifies domain polarization parallel and perpendicular to the surface and is compared with first principles Density Functional Theory (DFT) calculations.

In the tetragonal (P4mm) phase of BTO there are two possible FE distortions along the c-axis, parallel and anti-parallel to $\left[001\right]$. The polarization may also be along one of the four equivalent in-plane directions, identified as $\mathrm{P}^\mathrm{in}$. Out-of-plane polarization gives fixed positive or negative surface charge, whereas no net surface charge is expected for $\mathrm{P}^\mathrm{in}$.

The experiments were done with the NanoESCA photoemission spectrometer at the Elettra synchrotron.~\cite{Kroemker2008, Schneider2012} The stoichiometric BTO(001) single crystal sample (from SurfaceNet GmbH) was ozone cleaned ex-situ for 10 minutes immediately before introduction into the ultra-high vacuum (UHV) system.~\cite{Amy2004} The base pressure was better than $2\!\times\!10^{-8}$Pa. In the reciprocal space imaging mode~\cite{Kroemker2008, Mathieu2011}, which we call k-PEEM, the photoelectrons are detected with a wave vector resolution of about 0.05 $\mathrm{\AA}^\mathrm{-1}$ for all k$_\mathrm{parallel}$ and an energy resolution of 200 meV.

\begin{figure}
  \includegraphics[scale=0.6]{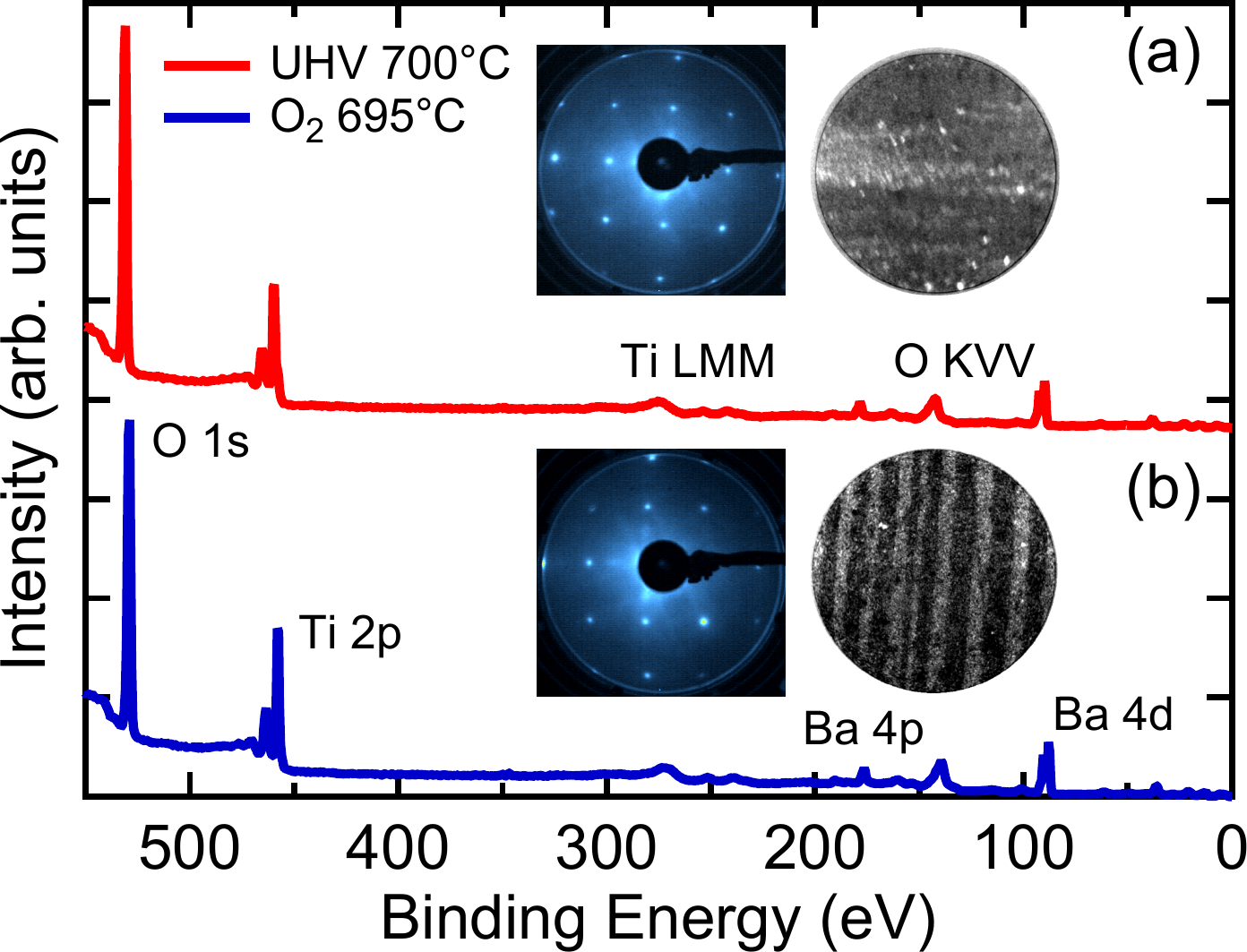}
  \caption{XPS survey spectra of BTO(001) after (a) annealing in UHV at 700$\degC$ and (b) annealing in $6\!\times\!10^{-6}$ mbar of oxygen at 700$\degC$. The insets show the corresponding LEED (left) and PEEM (right) images. The PEEM field of view is 65$\mu$m.}
  \label{fig:Carac}
\end{figure}

The surface was prepared by 3 cycles of low energy Ar ion sputtering (500 eV) and annealing (650$^\mathrm{\circ}$C) in UHV. The final annealing step was done at 700$^\mathrm{\circ}$C for 30 minutes creating a clean, ordered surface with oxygen vacancies.~\cite{Hirata1994, Kuwabara1997} Then the sample was annealed at 695$^\mathrm{\circ}$C for 2 hours in an oxygen partial pressure of $6\!\times\!10^{-6}$ mbar to partially recover the oxygen stoichiometry. Figure~\ref{fig:Carac} shows the XPS survey acquired using a photon energy of 650 eV after UHV (red, upper curve) and oxygen (blue, lower curve) annealing. They are similar and free of carbon. The insets show the low energy electron diffraction (LEED) patterns; in both cases we observe sharp (1$\times$1) patterns. However, the threshold PEEM images in the right hand insets are very different. After UHV annealing no contrast is observed, whereas after oxygen annealing typical ordered stripe domains appear oriented  parallel to $\left[001\right]$ or $\left[010\right]$, providing evidence for an uniaxial in-plane strain.~\cite{Hoefer2012, Barrett2013} The contrast results from the different surface charge as a function of domain polarization.

\begin{figure}
  \centering
 \includegraphics[scale=0.6]{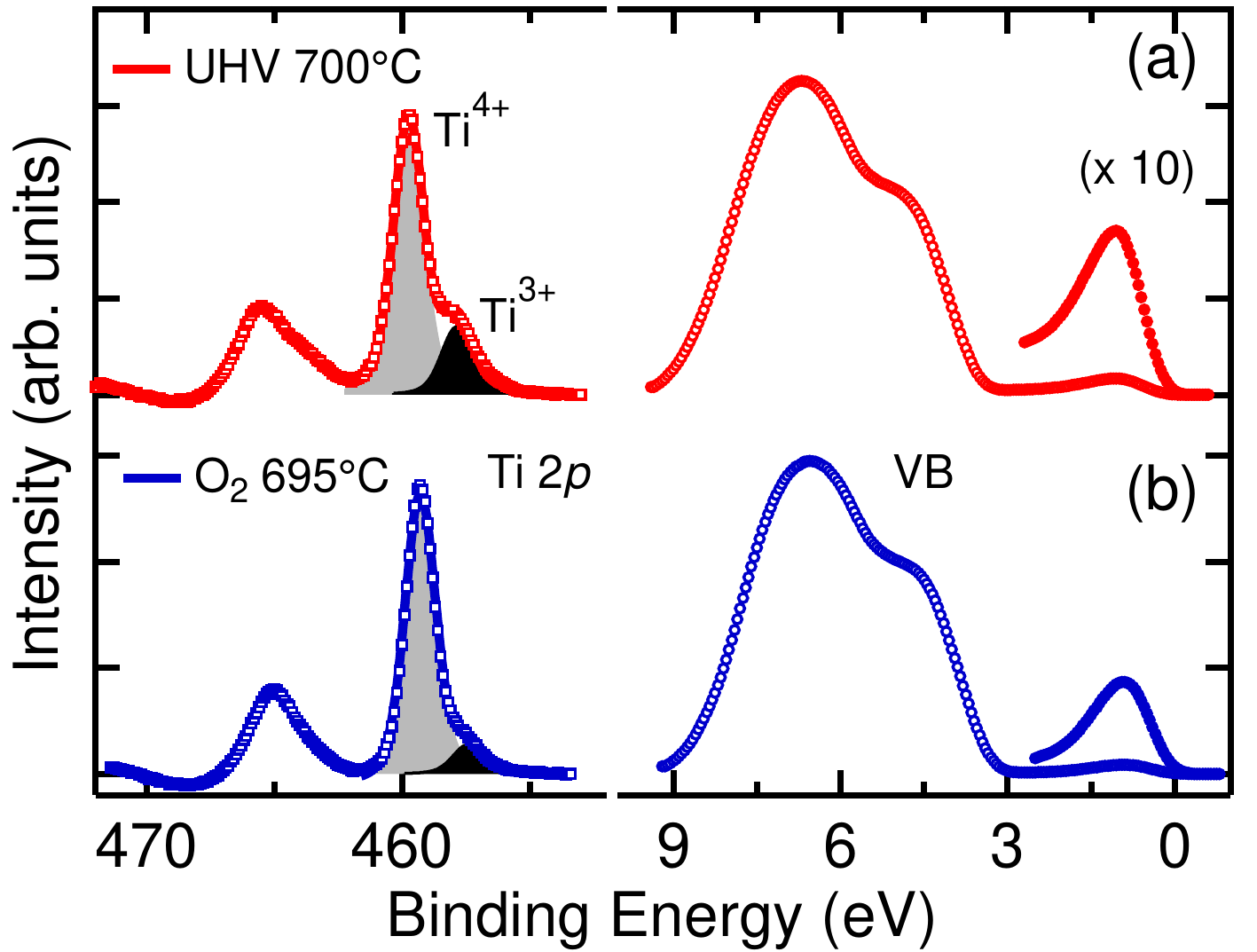}
  \caption{High resolution Ti 2\textit{p} core level (left) and valence band (right) spectra after UHV annealing (top) and annealing in oxygen (bottom).}
  \label{fig:VB_Ti2p}
\end{figure}

The Ti 2\textit{p} core level and valence band spectra using photon energies of 650 and 50 eV, respectively, are shown in Fig.~\ref{fig:VB_Ti2p}. The low binding energy Ti 2p component (black) is associated with Ti$^\mathrm{3+}$ ions, correlated to the appearance of states in the band gap, approximately 1 eV below the Fermi level shown on the right hand side of Fig.~\ref{fig:VB_Ti2p}. From the relative intensities of the Ti$^\mathrm{4+}$ and Ti$^\mathrm{3+}$ we estimate that the free carrier doping after annealing in UHV is 0.2 electrons per unit cell. This decreases to 0.09 e/u.c. after oxygen annealing. The first figure is higher than the predicted critical doping for ferroelectric displacements~\cite{Iwazaki2012, Wang2012} and explains the absence of domain contast in PEEM after UHV annealing. After oxygen annealing, the doping is lower than the critical value and ferroelectric stability is recovered, evidenced by the appearance of stripe domain contrast in the PEEM images. XPS measurements \footnote{See Supplemental Material at [URL].} at 900 eV (more bulk sensitive) confirm that the vacancy concentration is constant over the first few nanometers. The O 1\textit{s} and Ba 4\textit{d} spectra are identical for both photon energies~\cite{Note1}, showing that no phase separation or surface segregation has taken place. Therefore, the main result of UHV and oxygen annealing is to modify the doping level.

Based on this result and in the following analysis we suggest that the main effect of UHV annealing is to increase the oxygen vacancy concentration and thus the electron doping, creating a more metallic-like surface. The oxygen desorption energy is much lower at the surface than in the bulk ~\cite{Wang2012b}. Conversely, oxygen annealing reduces the electron doping allowing the surface to support the ferroelectric state.

\begin{figure}
  \centering
  \includegraphics[scale=0.5]{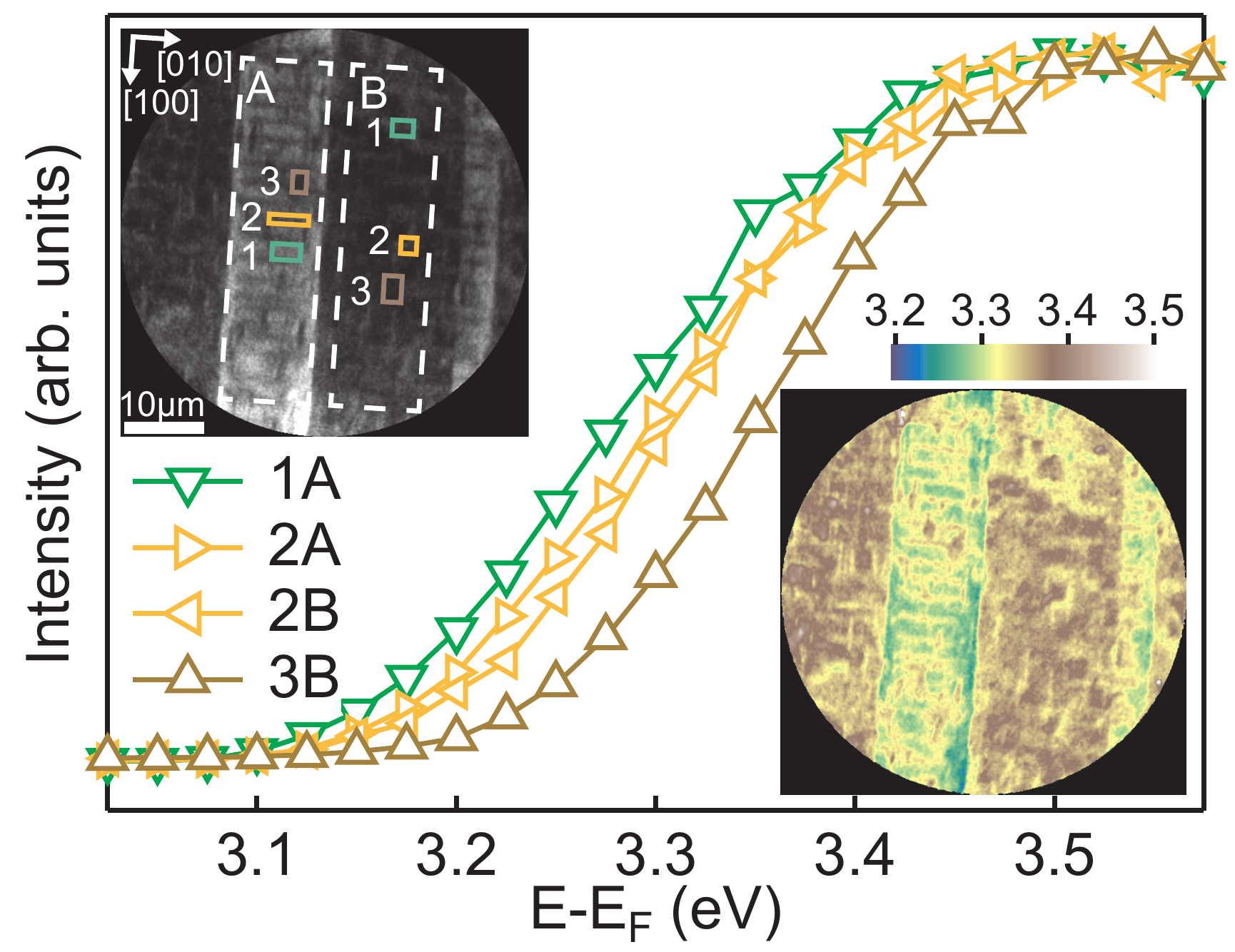}
  \caption{Local spectra extracted from rectangles 1A, 2A, 2B and 3B in upper image. Typical threshold PEEM image (upper inset) after annealing in oxygen with two main ROIs, A and B, each with finer domain structure. Threshold map (lower inset) constructed from a pixel by pixel error function fit to the image series in the full field of view. Main threshold values in A are 3.29 and 3.32 eV, in B 3.32 and 3.36 eV.}
  \label{fig:WF_PEEM}
\end{figure}

The upper inset of Fig.~\ref{fig:WF_PEEM} shows a typical PEEM image of another region of the same sample surface after oxygen annealing. The triple contrast indicates the presence of P$^\mathrm{\pm}$  and P$^\mathrm{in}$ polarization states. Regions of interest (ROIs), marked A and B, are selected corresponding to 10$\mu$m broad stripes. Both contain fine structure, defined by the rectangles 1A, 2A, 2B and 3B, indicating a high degree of sub-micron domain ordering. The local threshold spectra from this fine structure have different photoemission thresholds. The two lower energy spectra (1A and 2A) are in ROI A whereas ROI B contains the two higher energy spectra (2B and 3B). 2A and 2B are virtually identical, suggesting they are due to the same polarization. From consideration of surface charge they must be associated with P$^\mathrm{in}$. The lower inset of Fig.~\ref{fig:WF_PEEM} is the threshold map generated from a pixel by pixel error function fit~\cite{Mathieu2011} across the entire field of view of the image series. Three distinct values are obtained, 3.29, 3.32 and 3.36 eV with a standard deviation of 14 meV.~\cite{Note1} The intermediate value (3.32 eV), represented by the lighter (yellow) regions, is present in both ROIs. In ROI A the two low threshold values (color coded green and yellow) are dominant but there is a small proportion of high threshold values (color code brown). Thus in ROI A there are indeed three polarizations, P$^\mathrm{+}$, P$^\mathrm{-}$ and P$^\mathrm{in}$, but the proportion of P$^\mathrm{-}$ is much higher than P$^\mathrm{+}$. The opposite is true of ROI B.

\begin{figure}
  \centering
  \includegraphics[width=7cm,clip]{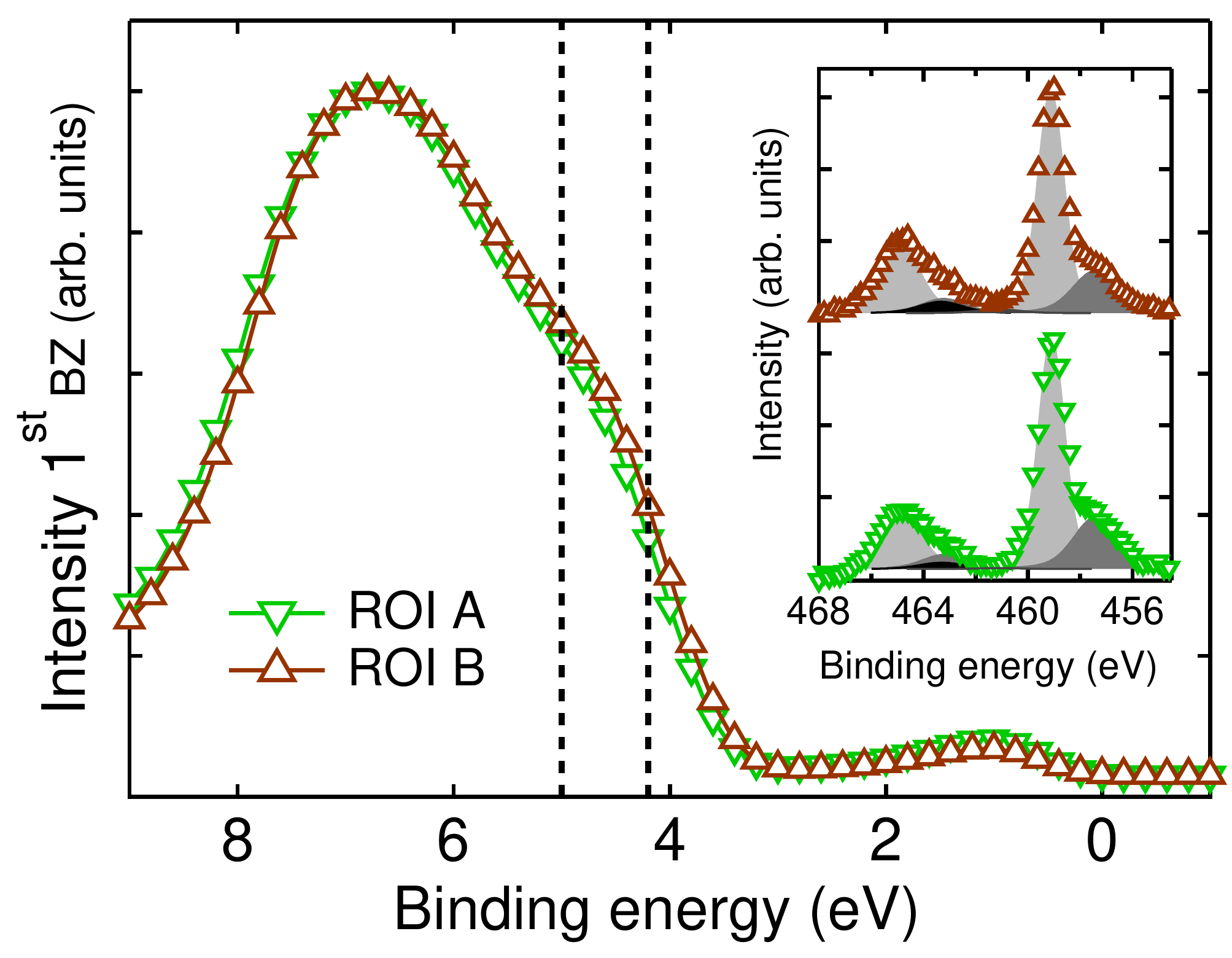}
  \caption{Micro-spectra obtained by placing a 5 $\mu$m iris in an intermediate image plane to select ROIs A (green downwards triangles) and B (brown, upwards triangles). Main figure shows the valence band spectra, the inset the Ti 2\textit{p} spectra. For clarity only 1 out of 5 (4) experimental points are shown for the valence band (Ti 2\textit{p}) spectrum. The vertical dotted lines indicate the positions of the constant energy cuts in the experimental band structure in Fig. ~\ref{fig:kPEEM_cuts}.}
  \label{fig:MicroSpectra}
\end{figure}

\begin{figure}
  \centering
  \includegraphics[width=8.5cm,clip]{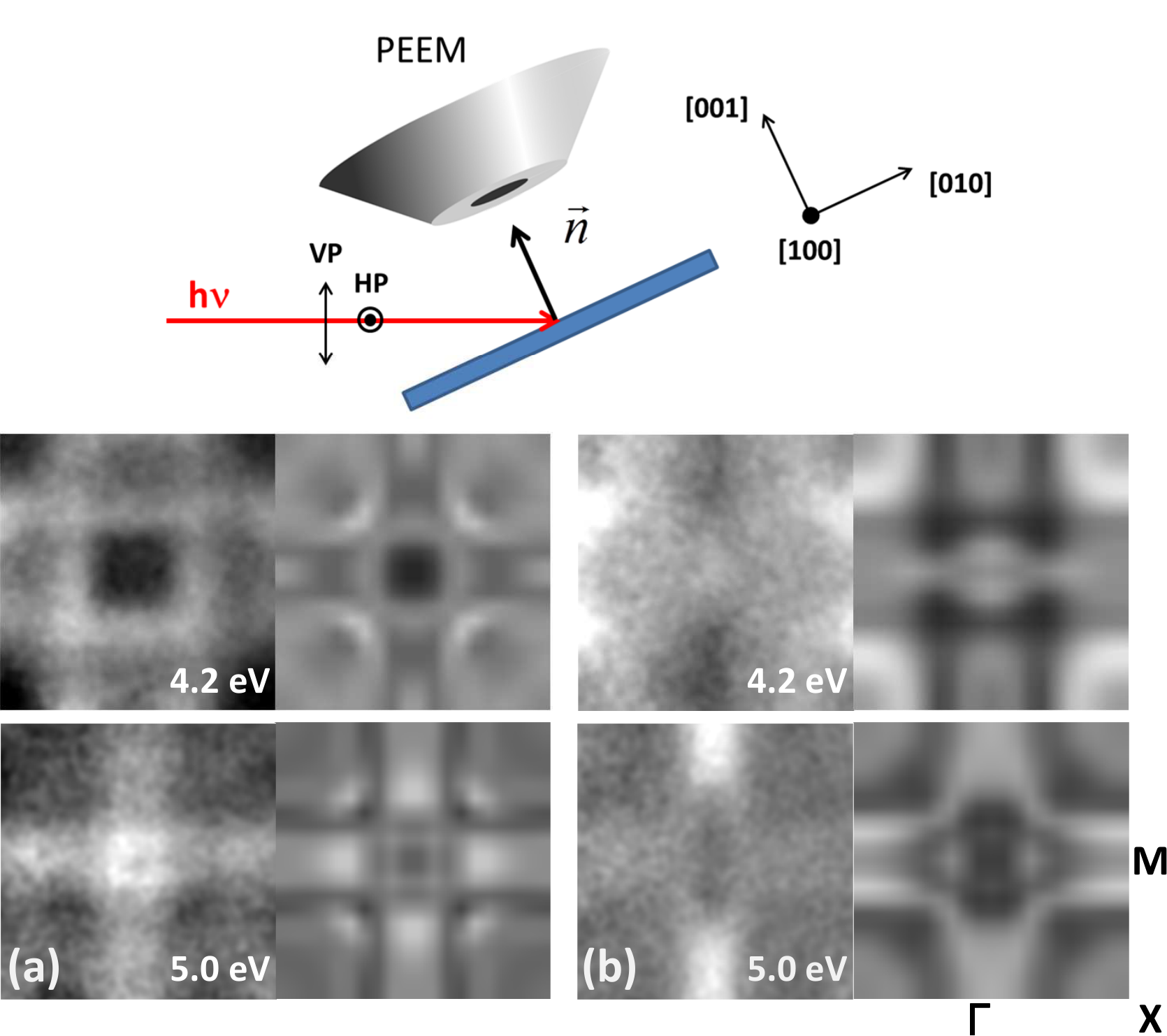}
  \caption{Top: schematic of the experimental geometry showing the direction of the linearly polarized light with respect to the crystallographic orientation. Main: constant energy cuts in the 1st Brillouin zone at 4.2 (upper) and 5.0 eV (lower) of the k-PEEM data from ROI A (B is similar~\cite{Note1}) using (a) VP and (b) HP light. The right hand panels show the theoretical calculations described below. $\left[001\right]$ is perpendicular to the paper and $\left[100\right]$ is horizontal.}
  \label{fig:kPEEM_cuts}
\end{figure}

Using an iris aperture of approximately 5 $\mu$m diameter we have carried out micro-XPS in the ROIs A and B. The valence band and (inset) Ti 2\textit{p} core level spectra are presented in Fig.~\ref{fig:MicroSpectra}. The micro-spectra extracted from ROIs A and B reflect the average of the in-plane and out-of-plane polarization in each zone resulting in their small relative shift (50 meV). The vertical dotted lines indicate the positions of the constant energy cuts in the band structure presented in Fig.~\ref{fig:kPEEM_cuts}. Interestingly, the oxygen vacancy concentration, as deduced from the Ti$^\mathrm{3+}$ component, is 13\% higher in ROI A (green, lower curve) corresponding to an effective doping of 0.1 e/u.c. than in ROI B (0.08 e/u.c.), suggesting a polarization dependence of the oxygen vacancy concentration.~\cite{Mi2012} In comparison, DFT-based calculations in the presence of oxygen vacancies predict a critical electron density of 0.06 e/u.c.~\cite{Iwazaki2012} Our evaluation of the doping level is based on the assumption that the Ti$^\mathrm{4+}$ state is reduced to Ti$^\mathrm{3+}$ but it is unlikely that a simple charge transfer model is sufficient to describe the electron density changes. The carriers released by the defect (i) may remain slightly bonded in the form of a polaron and (ii) keep a rather localized character due to the correlated nature of the Ti 3\textit{d} orbital.~\cite{Shimada2013} However, the general trend and the presence of a paraelectric-ferroelectric structural transition in a metallic-like state agree with the theoretical predictions.

We used linear dichroism in photoemission to investigate further the ferroelectric state in both ROIs. Linear dichroism in photoelectron spectroscopy~\cite{Smith1976} can be used to study both orbital~\cite{Ouardi2011} and structural symmetry~\cite{Cherepkov1993}. It should therefore be sensitive to orthogonal FE polarization directions.

The geometry is shown in the schematic in Fig.~\ref{fig:kPEEM_cuts}. The soft X-rays (50 eV) are incident at 25$^\circ$ with respect to the surface. The k-PEEM data are acquired from 1 eV above to 9 eV below the Fermi level in 50 meV steps (9 seconds per image). The resulting 2nd derivative data stack $I(E, k_x, k_y)$ gives the full 2D band dispersion parallel to the surface. We have used vertically (VP) and horizontally (HP) polarized light. For VP the light polarization on the sample is 60\% \textit{p}-polarized, with 40\% \textit{s}-polarized along $\left[010\right]$. For HP, the light is 100\% \textit{s}-polarized along $\left[100\right]$. From the dipole matrix elements we would expect \textit{p}-polarized light to be sensitive to distortions perpendicular to the surface plane, i.e. parallel or antiparallel to $\left[001\right]$ whereas \textit{s}-polarized light should be sensitive to in-plane distortions. The \textit{s}-polarized light should also enhance the emission intensity from the \textit{d}-orbitals.~\cite{Ouardi2011} The photon energy is such that k$_\mathrm{perp}$ is at a $\Gamma$ point (k$_\mathrm{perp}$ = 2$\pi$/a (0,0,2)).~\cite{Takizawa2009}

Figure~\ref{fig:kPEEM_cuts} shows constant energy cuts in the 1st Brillouin zone at 4.2 (upper) and 5.0 eV (lower) of the k-PEEM data from ROI A (B is similar~\cite{Note1}) using (a) VP and (b) HP light. The right hand panels show the theoretical calculations described below. There are important differences between VP and HP light. The band structure using VP light always has fourfold symmetry whereas for HP the symmetry is twofold. The fourfold symmetry with VP confirms the presence of out-of-plane polarization. Apart from the opposite surface charge the dispersion relations should be similar for both P$^\mathrm{+}$ and P$^\mathrm{-}$. The HP cuts show that there is preferential in-plane polarization, parallel and antiparallel to $\left[100\right]$, rather than in all four P$^\mathrm{in}$ directions. One would also expect evidence of twofold symmetry in the band structure using VP light. However, the 40\% \textit{s}-polarization is \textit{orthogonal} to that in HP, pointing to negligible in-plane tetragonal distortion  along $\left[010\right]$. This is strong evidence that the in-plane polarization are mainly along $\left[100\right]$ and $\left[-100\right]$.

With reference to the PEEM images in Fig.~\ref{fig:Carac} and Fig.~\ref{fig:WF_PEEM}, the domains in the as-received sample have grown along $\left[100\right]$ or $\left[010\right]$ and not, for example, along $\left[110\right]$ as observed by Schilling et al\cite{Schilling2011}. This must be due to elastic energy, however, in our experiment we do not have sufficient control of the strain imposed by the sample holder in order to make a quantitative correlation.

Density-functional calculations for in and out-of-plane polarizations in BTO were performed using the SIESTA code~\cite{Ordejon1996, Soler2002}. We use the local density approximation (LDA), Troullier-Martins pseudo-potentials and a basis of numerical atomic orbitals, extended up to triple $\zeta$ for Ti 3\textit{d} and 4\textit{s}, and O 2\textit{p}. The real space grid is determined by a 400 Ryd cut-off and the range of atomic orbitals by a 0.001 Ryd energy shift. Nine-layer, (001)-oriented slabs with TiO$_\mathrm{2}$ terminations are used. The geometry is relaxed using a conjugate-gradient algorithm until all the Cartesian components of the atomic forces are less than 0.04 eV/$\mathrm{\AA}$. Two configurations are considered: (i) an in-plane polarization (a = 3.94 $\mathrm{\AA}$) and (ii) an out-of-plane polarization (a = 3.995 $\mathrm{\AA}$), stabilized by superimposing an external electric field (E$_\mathrm{ext}$=1.2 V/$\mathrm{\AA}$) perpendicular to the surface. This is necessary to screen the depolarizing field and is associated to a saw-tooth electrostatic potential with a discontinuity in the middle of the vacuum layer (20 $\mathrm{\AA}$) to preserve the periodicity of the potential. A (1$\times$1) surface unit cell was assumed, as observed in LEED. A 6$\times$6$\times$1 sampling of the Brillouin Zone was used, refined up to 20$\times$20$\times$1 to accurately compute the band structure in the relaxed geometry.

The LDA results are shown as insets in Fig.~\ref{fig:kPEEM_cuts}. To compare with experiment they have been Gaussian broadened by 0.3 eV. It should also be borne in mind that any comparison between DFT based band structure and experiment is limited since we measure a spectral function which can show significant differences with respect to the calculated Kohn-Sham band structure~\cite{Chambers2004, Damascelli2004}. Within these limits we see that the agreement is rather good, the symmetry and the main band structures are reproduced. It is remarkable that despite the low contrast in real space PEEM, reciprocal space imaging is capable of identifying the surface polarization present in both zones, providing additional evidence for the ferroelectric state below critical doping.~\cite{Wang2012}

Direct imaging of the band structure from domains using standard ARPES is impossible because of the beam size. One exciting application of the method presented here is the quantification of the dispersion relations at the surface of thin films with a view to identifying bands which may hybridize with a magnetic overlayer and give rise to magneto-electric coupling.

In summary, using a combination of in-situ UHV and oxygen annealing we can control the surface doping of BaTiO$_\mathrm{3}$(001) and the paraelectric-ferroelectric transition. For doping levels below the critical value, ferroelectric stability leads to domain formation with in and out-of plane polarizations reflected in the band structure symmetry. The results compare well with first principles calculations and represent a breakthrough in the study of the electronic structure of ferroelectric surfaces. The unambiguous observation of domain ordering means that the link between band structure and polarization could be generalized to undoped ferroelectric systems.

\begin{acknowledgments}
J.R. and J. D. are funded by  CEA Ph.D. grants. We acknowledge Elettra for provision of synchrotron radiation facilities.
\end{acknowledgments}

\bibliography{./kPEEM_BTO}

\end{document}